\documentclass[preprint, pt1]{elsarticle}

\usepackage[latin1]{inputenc}                    
\usepackage{graphicx}                            
\usepackage{latexsym}                            
\usepackage{amsfonts}                            
\usepackage{amssymb}                             
\usepackage{amsmath}                             
\usepackage[mathscr]{eucal}                      
\usepackage{dcolumn}                             
\usepackage{theorem}                             
\usepackage{footnote}
\usepackage{enumitem}
\usepackage{bm}
\usepackage{color}
\usepackage{natbib}
\usepackage[margin=2.5cm]{geometry}%
\usepackage{placeins}

\def\be{\begin{equation}}
\def\ee{\end{equation}}
\def\bea{\begin{eqnarray}}
\def\eea{\end{eqnarray}}

\def\gg{\gamma \gamma}
\def\gZ{\gamma Z}

\def\qwe{$Q_{\text{weak}} \,\,$}
\def\mo{MOLLER }

\def\bgZv{$\square_{\gZ}^V \,$}

\def\regzv{$\Re e \, \square_{\gZ}^V$}		

\def\fogg{F_1^{\gg}}
\def\flgg{F_L^{\gg}}

\def\ftgz{F_2^{\gZ}}
\def\fogz{F_1^{\gZ}}
\def\flgz{F_L^{\gZ}}

\def\g2{GeV$^2$}
\def\ztl{$\kappa_C^{T, L}$ }

\begin{document}

\title{Quark-hadron duality constraints on $\gamma Z$ box corrections \\
	to parity-violating elastic scattering}

\author[man]{N.~L.~Hall }
\author[man]{P.~G.~Blunden }
\author[jlb]{W.~Melnitchouk }
\author[adl]{A.~W.~Thomas }
\author[adl]{R.~D.~Young }

\address[man]{\mbox{Department of Physics and Astronomy,
	University of Manitoba}, Winnipeg, MB, Canada R3T 2N2}

\address[jlb]{\mbox{Jefferson Lab, 
	12000 Jefferson Avenue, Newport News, Virginia 23606, USA}}

\address[adl]{\mbox{ARC Centre of Excellence for Particle Physics 
	at the Terascale and CSSM}, Department of Physics,	\\
	University of Adelaide, Adelaide SA 5005, Australia}

\date{\today}

\begin{abstract}
We examine the interference $\gZ$ box corrections to parity-violating
elastic electron--proton scattering in the light of the recent
observation of quark-hadron duality in parity-violating deep-inelastic
scattering from the deuteron, and the approximate isospin independence
of duality in the electromagnetic nucleon structure functions down to
$Q^2 \approx 1$~GeV$^2$.  Assuming that a similar behavior also holds
for the $\gZ$ proton structure functions, we find that duality
constrains the $\gZ$ box correction to the proton's weak charge to be
$\Re e\, \square_{\gZ}^V = (5.4 \pm 0.4) \times 10^{-3}$
at the kinematics of the \qwe experiment.
Within the same model we also provide estimates of the $\gZ$ corrections
for future parity-violating experiments, such as MOLLER at Jefferson Lab
and MESA at Mainz.
\end{abstract}

\begin{keyword}
Parity violation, proton weak charge, quark-hadron duality
\end{keyword}

\maketitle

\section{Introduction}
\label{sec:Intro}

Parity-violating precision measurements have for many years provided
crucial low-energy tests of the Standard Model.  Early efforts such as
the E122 experiment at SLAC \cite{Prescott1978tm, Prescott1979dh}
firmly established the SU(2)$\times$U(1) model as the theory of the
unified electroweak interactions.  Modern-day experiments use parity
violation to probe physics beyond the Standard Model.  One of the most
recent parity-violating measurements is the \qwe experiment at
Jefferson Lab \cite{Armstrong2012ps}, which aims to measure the
proton's weak charge to 4\% accuracy.  With an initial analysis
of a subset of the data already reported \cite{Androic2013},
the analysis of the full data set is expected in the near future.

For the precision requirements of the \qwe experiment, the weak charge 
of the proton, defined at tree level as
$Q_W^p = 1 - 4 \sin^2 \theta_W$,
must also include radiative corrections.
Including these corrections at the 1-loop level, the weak charge can be
written as \cite{Erler2003yk}
\bea
Q_W^p &=& \left( 1 +\Delta\rho + \Delta_e \right)
	  \left( 1 - 4 \sin^2\theta_W(0) + \Delta_e^{'} \right)
       + \square_{WW} + \square_{ZZ} + \square_{\gZ}(0), 
\label{eq:qwHO}
\eea
where $\sin^2\theta_W(0)$ is the weak mixing angle at zero momentum
transfer, and the electroweak vertex and neutral current correction
terms $\Delta \rho$, $\Delta_e$ and $\Delta_e'$ have been calculated
to the necessary levels of precision \cite{Erler2003yk}.
The weak box corrections $\square_{WW}$ and $\square_{ZZ}$ are
dominated by short-distance effects and can also be computed
perturbatively to the required accuracy.

On the other hand, the final term in Eq.~\eqref{eq:qwHO}, the $\gZ$
box contribution, depends on both short- and long-distance physics
and therefore requires nonperturbative input.  Considerable attention
has been given to the analysis of this term, for both the vector
electron--axial vector hadron coupling to the $Z$, $\square_{\gZ}^A$
(which is relevant for atomic parity violation experiments)
\cite{Marciano1983, Marciano1984, Blunden2011rd, Blunden2012ty},
and the axial electron--vector hadron coupling, $\square_{\gZ}^V$
(which because of its strong energy dependence makes important
contributions to the \qwe experiment) \cite{Gorchtein2008px,
Sibirtsev2010zg, Rislow2010vi, Gorchtein2011mz, Hall2013hta}.
The most accurate technique to evaluate the latter is a dispersion
relation.  While constraints from parton distribution functions
(PDFs) and recent parity-violating deep-inelastic scattering (PVDIS)
data \cite{Wang2013kkc, Wang2014bba} provide a systematic way of
reducing the errors on this correction \cite{Hall2013hta}, some
uncertainty remains about the model dependence of the low-$Q^2$
input.

The E08-011 electron--deuteron PVDIS experiment at Jefferson Lab
not only allowed an accurate determination of the $C_{2q}$
electron--quark effective weak couplings \cite{Wang2014bba}, but
also presented the first direct evidence for quark-hadron duality
in $\gZ$ interference structure functions, which was verified at
the (10--15)\% level for $Q^2$ down to $\approx 1$~GeV$^2$
\cite{Wang2013kkc}.
In general, quark-hadron duality refers to the similarity of
low-energy hadronic cross sections, averaged over resonances,
with asymptotic cross sections, calculated at the parton level
and extrapolated to the resonance region.
It is manifested in many different hadronic observables
\cite{Melnitchouk2005} and was first observed in deep-inelastic
scattering (DIS) by Bloom and Gilman \cite{Bloom1970, Bloom1971}.
Subsequent studies have quantified the validity of duality for
various spin-averaged and spin-dependent electromagnetic structure
functions, as well as in neutrino scattering and for different
targets \cite{Niculescu2000, Niculescu2000a, Airapetian2003,
Arrington2003nt, Wesselmann2007, Psaker2008, Malace2009, Malace2010},
establishing the phenomenon as a general feature of the strong
interaction.

Furthermore, recent analysis of moments of the free neutron
electromagnetic structure function \cite{Niculescu2015} has
demonstrated that duality in the lowest three neutron moments
is violated at a similar level ($\lesssim 10\%$) as in the proton
for $Q^2 \geqslant 1$~GeV$^2$ \cite{Niculescu2000, Niculescu2000a,
Malace2009}.  This suggests that the isospin dependence of duality
and its violation is relatively weak.  It is reasonable therefore
to expect that duality may also hold to a similar degree for the
$\gZ$ structure functions, which are related to the electromagnetic
structure functions by isospin rotations.

In this paper we discuss the extent to which quark-hadron duality in
$\gZ$ structure functions can provide additional constraints on the
$\square_{\gZ}^V$ corrections, and in particular the contributions
from low hadronic final state masses $W$ and $Q^2 \sim 1$~GeV$^2$.
In Sec.~\ref{sec:DualSM} we illustrate the realization of duality
in the moments of the proton and neutron electromagnetic structure
functions using empirical parametrizations of data in the resonance
and DIS regions down to $Q^2 = 1$~GeV$^2$.
Motivated by the approximate isospin independence of duality in
electromagnetic scattering from the nucleon, in Sec.~\ref{sec:ImplBgZ}
we explore the consequences of duality in the $\gZ$ structure functions
for the energy dependence of the $\square_{\gZ}^V$ correction, and
especially the limits on its overall uncertainty.
Finally, in Sec.~\ref{sec:Con} we summarize our findings and discuss
their implications for the analysis of the \qwe experiment as well as
future parity-violating experiments such as MOLLER at Jefferson Lab
\cite{MOLLER} and MESA at Mainz \cite{MESA}.

\section{Duality in electromagnetic structure functions}
\label{sec:DualSM}

Historically, the observation of duality in inclusive electron
scattering \cite{Bloom1970, Bloom1971} predates the development of
QCD and was initially formulated in the language of finite-energy
sum rules.  Within QCD, duality was reinterpreted within the
operator product expansion through moments of structure functions
\cite{DeRujula1977}, with duality violations associated with matrix
elements of higher twist (HT) operators describing multi-parton physics.
The extent to which inclusive lepton--nucleon cross sections can be
described by incoherent scattering from individual partons through
leading twist (LT) PDFs can be quantified by studying the $Q^2$
dependence of the structure function moments.
At low $Q^2$, corrections to the LT results arise not only from
multi-parton processes, but also from kinematical target mass
corrections (TMCs), which, although $1/Q^2$ suppressed, arise from
LT operators.
To isolate the genuine duality-violating HT effects, one can consider
Nachtmann moments of structure functions \cite{Nachtmann1973}, which
are constructed to explicitly remove the effects of higher spin
operators and the resulting TMCs.

Specifically, the Nachtmann moments of the $F_1$ and $F_2$ structure
functions are defined as \cite{Nachtmann1974, Nachtmann_note}
\bea
\mu_1^{(n)}(Q^2)
&=& \int_0^1 dx \, \frac{\xi^{n+1}}{x^3}
    \left[ x F_1(x,Q^2) + \frac{1}{2}\rho^2 \eta_n F_2(x,Q^2)
    \right],
\label{eq:mu1}			\\
\mu_2^{(n)}(Q^2)
&=& \int_0^1 dx \, \frac{\xi^{n+1}}{x^3}\,
    \rho^2 (1 + 3\eta_n) F_2(x,Q^2),
\label{eq:mu2}
\eea
where
\be 
\xi = \frac{2 x}{1 + \rho}
\label{eq:xi}
\ee
is the Nachtmann scaling variable \cite{Nachtmann1974, Greenberg1971},
with $x = Q^2/(W^2 - M^2 + Q^2)$ the Bjorken scaling variable,
$\rho^2 = 1 + 4 M^2 x^2/Q^2$, and $M$ the nucleon mass.
The variable $\eta_n$ is given by
\bea
\eta_n
&=& \frac{\rho-1}{\rho^2}
    \left[ \frac{n + 1 - (\rho+1)(n + 2)}{(n+2)(n+3)} \right],
\label{eq:eta_n}
\eea
and vanishes in the $Q^2 \to \infty$ limit.
In that limit the moments $\mu_i^{(n)}$ approach the standard
Cornwall-Norton moments \cite{Cornwall1969},
\bea
\mu_i^{(n)}(Q^2)
&\longrightarrow& M_i^{(n)}(Q^2)\,
 =\, \int_0^1 dx\, x^{n-i}\, F_i(x,Q^2),\ \ \ \ i=1,2.
\label{eq:CNmom}
\eea
At finite $Q^2$, while the $\mu_2^{(n)}$ moments depend only on the
$F_2$ structure function, the $\mu_1^{(n)}$ moments have contributions
from both the $F_1$ and $F_2$ structure functions.  Because the latter
contribution is proportional to $\eta_n$, it vanishes at large $Q^2$,
so that the $\mu_1^{(n)}$ moments are generally dominated by the
$F_1$ structure function at large $Q^2$.

\begin{figure}[t]
\begin{center}
\includegraphics[width=1.05\textwidth]{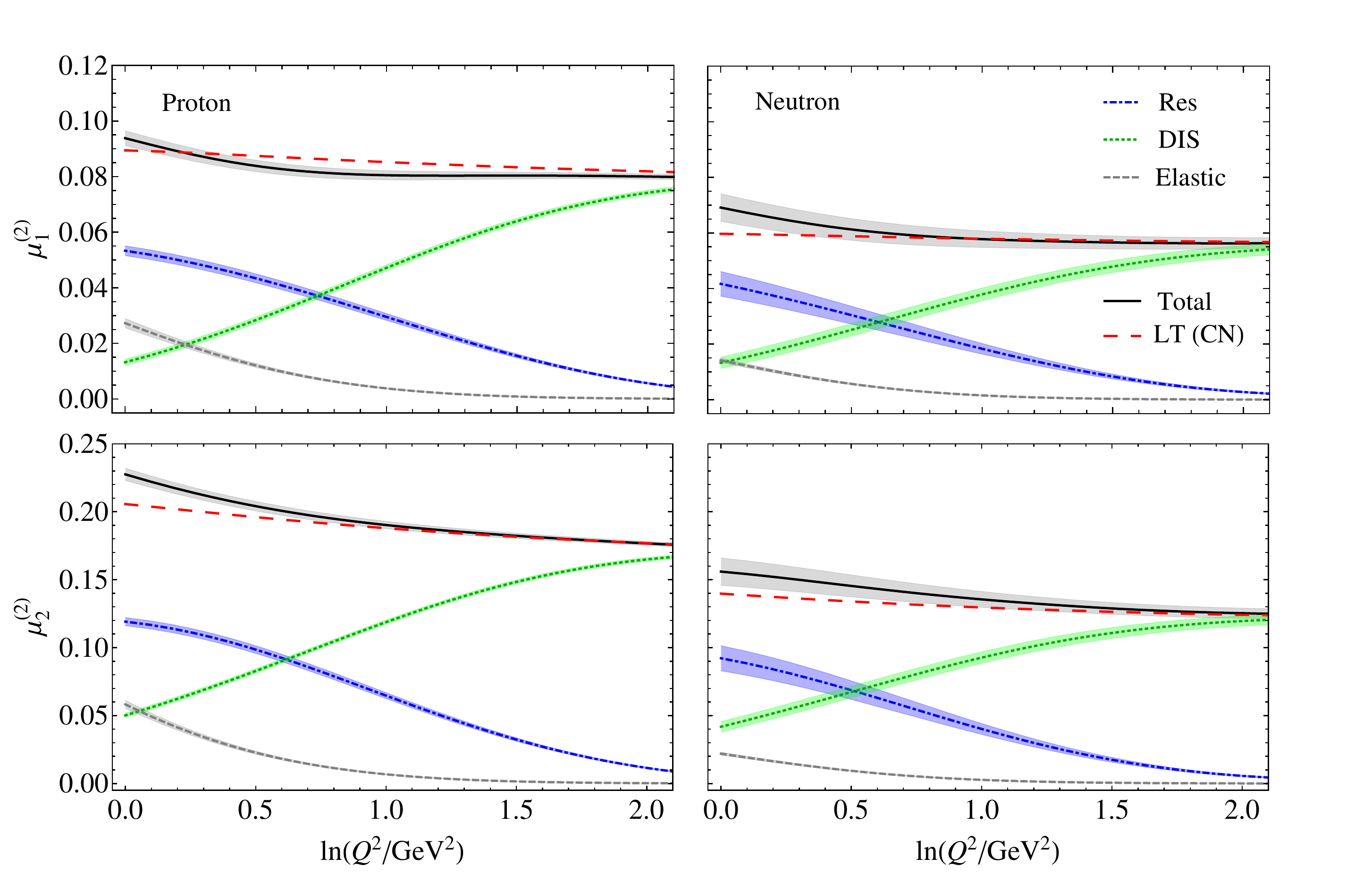}
\caption{The proton (left panels) and neutron (right panels)
	electromagnetic	$F_1^{\gg}$ (top) and $F_2^{\gg}$ (bottom)
	structure function moments.  The total Nachtmann moments
	(black solid lines) include contributions from the
	resonance ($W^2 \leqslant 6$~GeV$^2$, blue dot-dashed lines)
	and DIS ($W^2 > 6$~GeV$^2$, green dotted lines) regions,
	as well as the elastic contributions (gray dashed lines),
	and are compared with the Cornwall-Norton moments of
	the LT structure functions (red long-dashed lines).}
\label{fig:mugg}
\end{center}
\end{figure}

Duality in unpolarized electron--nucleon scattering has been studied
most extensively for the electromagnetic $F_2$ structure function
\cite{Niculescu2000, Niculescu2000a, Malace2009}, and to a lesser
extent for the $F_1$ (or longitudinal $F_L$) structure function
\cite{Melnitchouk2005, Monaghan2013}.
The latter is generally more difficult to access experimentally,
as it requires precise longitudinal--transverse separated cross
section measurements, or equivalently the $\sigma_L/\sigma_T$
cross section ratio.
In Fig.~\ref{fig:mugg} the workings of duality in the $n=2$
Nachtmann moments of the proton and neutron $F_1^{\gg}$ and
$F_2^{\gg}$ structure functions are illustrated over the range
$1 \leqslant Q^2 \leqslant 8$~GeV$^2$.
For the low-$W^2$ contributions, $W^2 \leqslant 6$~GeV$^2$, the
resonance-based fit to the electromagnetic structure function
data from Christy and Bosted \cite{Christy2010} is used.
For the DIS region at higher $W^2$ values, $W^2 > 6$~GeV$^2$,
this is supplemented by the ABM global QCD fit \cite{Alekhin2012}
to high-energy data, which includes LT, TMC and HT contributions.
Since LT evolution is logarithmic in $Q^2$, at large $Q^2$
the moments are predicted to become flat in $\ln Q^2$.
While the individual resonance and DIS region contributions,
as well as the elastic ($W=M$) component, are strongly $Q^2$
dependent in the region of low $Q^2$ shown in Fig.~\ref{fig:mugg},
remarkably their sum exhibits only very mild $Q^2$ dependence
down to $Q^2 \approx 1$~GeV$^2$.
This is the classic manifestation of duality observed by Bloom and
Gilman \cite{Bloom1970, Bloom1971}, in which the total empirical
moments resemble the LT contributions down to surprisingly low
momentum scales.
Note that since the Nachtmann moments are constructed to remove
higher spin operators that are responsible for TMCs, in the absence
of HTs one would expect the Nachtmann moments of the total structure
functions to equal the Cornwall-Norton moments of the LT functions,
$\mu_i^{(n)}({\rm LT+TMC}) = M_i^{(n)}({\rm LT)}$ \cite{Steffens2006}.

This expectation is clearly borne out in Fig.~\ref{fig:mugg},
where the total $\mu_1^{(2)}$ and $\mu_2^{(2)}$ moments are
very similar to the moments computed from the LT PDFs.
For the proton structure functions, the average violation of duality
in the range $1 \leqslant Q^2 \leqslant 2.5$~GeV$^2$ is 3\% and 4\%
for the $F_1^{\gg}$ and $F_2^{\gg}$ structure functions, respectively,
with the maximum violation being $\approx 5\%$ and $\approx 10\%$ at
the lower end of the $Q^2$ range.
For the neutron the maximum violation is slightly larger, with the
LT $F_1^{\gg}$ and $F_2^{\gg}$ moments being $\approx 14\%$ and
$\approx 10\%$ smaller than the full results, although the average
over this $Q^2$ range is 5\% and 8\%, respectively.
This is consistent with several previous phenomenological analyses
\cite{Virchaux1992, Alekhin2004, JR2014} of high-energy scattering
data which have found no indication of strong isospin dependence of HT
corrections.
%
Following Ref.~\cite{Christy2010}, we assign a 5\% error on the
proton $F_1^{\gg}$ and $F_2^{\gg}$ structure functions,
and a larger, 10\% error on the neutron structure function
\cite{Bosted2007xd}, reflecting the additional nuclear model
dependence in extracting the latter from deuterium data \cite{CJ13}.
For the elastic contribution a 5\% uncertainty is assumed for the
total elastic structure functions from Ref.~\cite{Kelly2004}.
For higher moments ($n > 2$), which are progressively more sensitive
to the high-$x$ (or low-$W$) region, the degree to which duality is
satisfied diminishes at lower $Q^2$ values \cite{Ji95}.

\section{Duality in $\gZ$ structure functions and implications
	for $Q_W^p$}
\label{sec:ImplBgZ}

In contrast to the electromagnetic structure functions which have been
studied extensively for many years, experimental information on the
interference $\gZ$ structure functions is for the most part nonexistent.
Some measurements of $F_2^{\gZ}$ and $xF_3^{\gZ}$ have been made at
very high $Q^2$ at HERA \cite{HERA}, where the $\gZ$ contribution
becomes comparable to the purely electromagnetic component of the
neutral current.  However, no direct measurements of $F_1^{\gZ}$ and
$F_2^{\gZ}$ for the proton exist in the $Q^2 \sim$~few~GeV$^2$ range
relevant for the evaluation of the $\gZ$ box correction to $Q_W^p$
\cite{Hall2013hta}.

In principle, the computation of the imaginary part of the
$\square_{\gZ}^V$ correction to the proton's weak charge at a given
incident energy $E$ requires knowledge of the $\gZ$ structure functions
over all kinematics,
\bea
\hspace{-1.0cm} 
\Im m\, \square_{\gZ}^V(E)
&=& \frac{1}{(s - M^2)^2}
    \int_{W_\pi^2}^s dW^2
    \int_0^{Q^2_{\rm max}} dQ^2\, \frac{\alpha(Q^2)}{1+Q^2/M_Z^2}
    \left[ \fogz
         + \frac{ s \left( Q^2_{\rm max}-Q^2 \right) }
		{ Q^2 \left( W^2 - M^2 + Q^2 \right) } \ftgz
    \right].
\label{eq:ImBoxV}
\eea
where $\alpha$ is the running electromagnetic coupling evaluated
at the scale $Q^2$, and $M_Z$ is the $Z$ boson mass.
The $W^2$ range covered in the integral lies between the inelastic
threshold, $W_{\pi}^2 = (M + m_\pi)^2$ and the total electron--proton
center of mass energy squared, $s = M^2 + 2 M E$, while the $Q^2$
integration range is from 0 up to $Q^2_{\rm max} = 2ME (1 - W^2/s)$.
(The small mass of the electron is neglected throughout.)
The real part of the $\gZ$ box correction which enters in
Eq.~(\ref{eq:qwHO}) can then be determined from the imaginary part
through an unsubtracted dispersion relation \cite{Gorchtein2008px,
Sibirtsev2010zg, Rislow2010vi, Gorchtein2011mz, Hall2013hta},
\bea
\Re e\, \square_{\gZ}^V (E)
&=& \frac{2E}{\pi}  {\cal P} \int_0^\infty dE' \frac{1}{E'^2-E^2}\,
    \Im m\, \square_{\gZ}^V(E'), 
\label{eq:DRv}
\eea
where ${\cal P}$ is the Cauchy principal value integral.
While the dispersion relation (\ref{eq:DRv}) is valid only for
forward scattering, because the \qwe experiment is performed at a
small scattering angle $\approx 6^\circ$, in practice it provides
a very good approximation.

Note that at high $Q^2$ and large $E$, the total correction
$\Re e\, \square_{\gZ}^V$ can also be expressed in terms of the
moments of the $F_1^{\gZ}$ and $F_2^{\gZ}$ structure functions by
switching the order of the integrations in Eqs.~(\ref{eq:ImBoxV})
and (\ref{eq:DRv}) and expanding the integrand in powers of
$x^2/Q^2$ \cite{Blunden2011rd}.
The higher order terms in $1/Q^2$ are then given in terms of
higher moments of the structure functions.  The expansion in
Ref.~\cite{Blunden2011rd} was performed in terms of the
Cornwall-Norton moments, but the expansion could also be
generalized to the Nachtmann moments in Eqs.~(\ref{eq:mu1})
and (\ref{eq:mu2}).
However, because this approximation neglects contributions from
the low-$W$ region, it is appropriate only for DIS kinematics
and is not directly applicable for the present application,
where the integrals are dominated by contributions at low $Q^2$
and $W^2$.
In particular, as we discuss below, at energy $E \sim 1$~GeV,
approximately 2/3 of the integral comes from the traditional
resonance region $W < 2$~GeV and $Q^2 < 1$~GeV$^2$.
In contrast, the contribution from the DIS region for $W > 2$~GeV
and $Q^2 > 1$~GeV$^2$ is $\approx 13\%$ at this energy.

In Refs.~\cite{Hall2013hta, Hall2013loa} the $F_1^{\gZ}$
and $F_2^{\gZ}$ structure functions were computed from the
phenomenological Adelaide-Jefferson Lab-Manitoba (AJM)
parametrization.
This is based on the electromagnetic structure functions described
in Sec.~\ref{sec:DualSM}, but appropriately rotated to the $\gZ$
case according to the specific $W^2$ and $Q^2$ region considered,
with the rotation parameters constrained by phenomenological PDFs
\cite{Hall2013hta} and recent PVDIS data \cite{Wang2013kkc,
Wang2014bba}.
In the AJM model the integrals over $W^2$ and $Q^2$ in
Eq.~(\ref{eq:ImBoxV}) are split into three distinct regions,
characterized by different physical mechanisms underlying the
scattering process.  In each region the most accurate parametrizations
or models of $F_1^{\gZ}$ and $F_2^{\gZ}$ available for the appropriate
kinematics are used.

\begin{figure}[t]
\begin{center}
\includegraphics[width=0.65\textwidth]{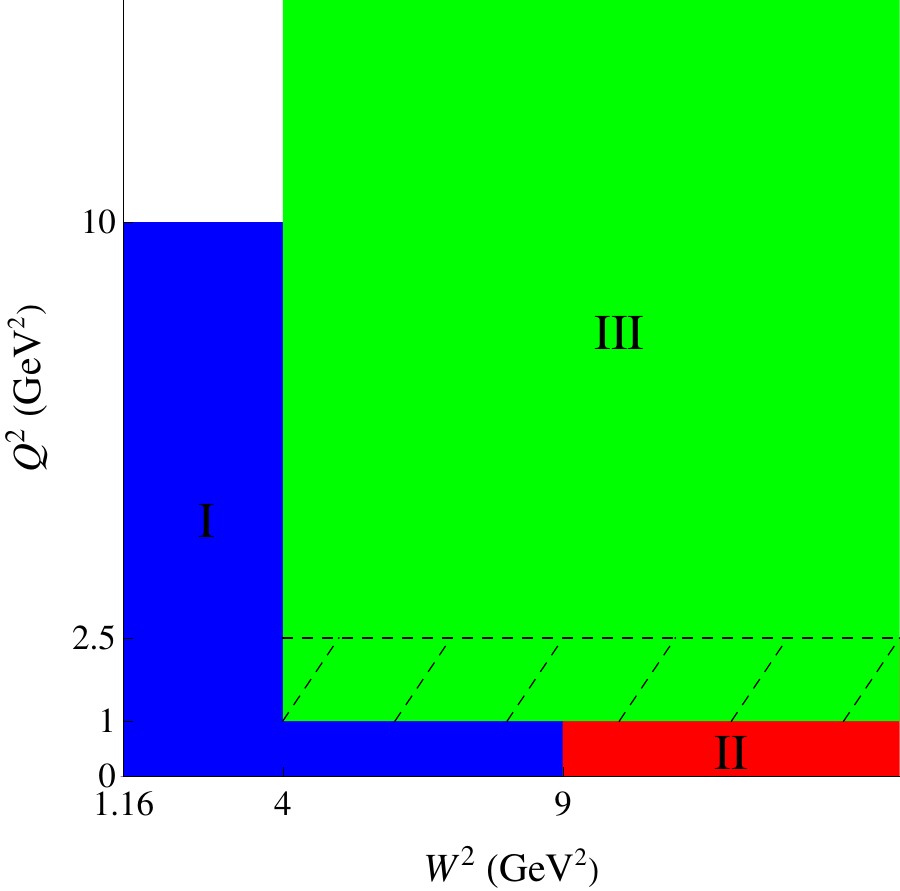}
\caption{Kinematic regions contributing to the
	$\square_{\gZ}^V$ integrals in the AJM model.
	Region I (blue) includes the nucleon resonance region
	  at low $W^2$ and $Q^2$;
	Region II (red) encompasses the low-$Q^2$, high-$W^2$
	  region described by Regge theory; and
	Region III (green) is the deep-inelastic region
	  characterized by LT PDFs.
	The shaded band between $Q^2=1$ and 2.5~GeV$^2$ represents
	the extension of Region III from its previous boundary in
	Ref.~\cite{Hall2013hta} ($Q^2=2.5$~GeV$^2$) to its current
	reach ($Q^2=1$~GeV$^2$).}
\label{fig:Q2W2}
\end{center}
\end{figure}

In the present analysis, we define the $W^2$ and $Q^2$ regions
as illustrated in Fig.~\ref{fig:Q2W2}.
``Region~I'' (low $Q^2$, low $W^2$) encompasses
  $0 \leqslant Q^2 \leqslant 10$~GeV$^2$
  for $W_\pi^2 \leqslant W^2 \leqslant 4$~GeV$^2$, and
  $0 \leqslant Q^2 \leqslant 1$~GeV$^2$
  for $4 < W^2 \leqslant 9$~GeV$^2$, using the $\gg \to \gZ$
  rotated Christy-Bosted parametrization \cite{Christy2010}
  of the resonance $+$ background structure functions.
For ``Region~II'' (low $Q^2$, high $W^2$), the vector meson
  dominance $+$ Regge model of Alwall and Ingelman
  \cite{Alwall2004wk} is used over the range
  $0 \leqslant Q^2 \leqslant 1$~GeV$^2$ and $W^2 > 9$~GeV$^2$.
Finally, for ``Region~III'' (high $Q^2$, high $W^2$) the
  perturbative QCD-based global fit from Alekhin {\it et al.}
  (ABM) \cite{Alekhin2012} is used for $Q^2 > 1$~GeV$^2$ and
  $W^2 > 4$~GeV$^2$, which includes LT as well as subleading
  $1/Q^2$ TMC and HT contributions.
For $x=1$, the elastic contributions to the structure functions
are computed using the form factor parametrizations from
Ref.~\cite{Kelly2004}.

While the uncertainties on the $\gZ$ structure functions in Region~III
are small --- typically a few \%, reflecting the errors on the PDFs
from which they are constructed through the simple replacement of
quark charges $e_q \to g_V^q$ --- the uncertainties in $F_1^{\gZ}$
and $F_2^{\gZ}$ are expected to be larger at lower $W^2$ and $Q^2$.
In the previous analyses of the $\gZ$ correction \cite{Hall2013hta,
Hall2013loa}, the PDF-based description was limited to
$Q^2 > 2.5$~GeV$^2$ (and $W^2 > 4$~GeV$^2$).
Motivated by the observation of duality in the proton and neutron
$F_1^{\gg}$ and $F_2^{\gg}$ structure functions, and in PVDIS from the
deuteron, as discussed in Sec.~\ref{sec:DualSM}, we further assume the
approximate validity of duality in the $\gZ$ proton structure functions
and extend the QCD description of Region~III down to $Q^2 = 1$~GeV$^2$.
Lowering the boundary of the DIS region, which is well constrained
by leading twist PDFs, to smaller $Q^2$ decreases the contribution
from Regions~I and II, and hence reduces the model uncertainty on the
$\gg \to \gZ$ rotation of the structure functions in this region.

Within the AJM $\gZ$ structure function parametrization, the most
uncertain elements are the \ztl continuum parameters used to relate
the high-mass, non-resonant continuum part of the $\gZ$ transverse
and longitudinal cross sections to the $\gg$ cross sections in the
generalized vector meson dominance model \cite{Alwall2004wk,
Sakurai1972wk}.
The \ztl parameters are fitted by matching the $\gZ$ to $\gg$
cross section ratios with the LT structure function ratios at
$Q^2 = 1$~GeV$^2$,
\bea 
\frac{\sigma_T^{\gZ} (\kappa_C^T)}{\sigma_T^{\gg}}
&=& \left. \frac{\fogz}{\fogg} \right|_{\rm LT},
\hspace*{1.5cm}
\frac{\sigma_L^{\gZ} (\kappa_C^L)}{\sigma_L^{\gg}}\
 =\ \left. \frac{\flgz}{\flgg} \right|_{\rm LT},
\label{eq:sigTFi}
\eea
where the longitudinal structure function $F_L$ is
related to the $F_1$ and $F_2$ structure functions by
$F_L = \rho^2 F_2 - 2x F_1$ \cite{Hall2013hta}.
(Note that, consistent with the duality hypothesis, we use the LT
structure functions in Region~III rather than the total structure
functions that may include the small subleading contributions
\cite{Alekhin2012}.)
The resulting fit values,
\be
\kappa_C^T\, =\, 0.36 \pm 0.15,  \qquad \qquad
\kappa_C^L\, =\, 1.5  \pm 3.1,
\label{eq:kappaC}
\ee
are obtained by averaging over the \ztl parameter determined from
10 fits with the ratios in Eq.~(\ref{eq:sigTFi}) matched at between
$W^2 = 4$~GeV$^2$ and 13~GeV$^2$.
These values are then used to compute the $\gZ$ structure functions
in the dispersion integral for $1 \leqslant Q^2 \leqslant 10$~GeV$^2$
and $W_\pi^2 \leqslant W^2 \leqslant 4$~GeV$^2$.
To allow for stronger violations of duality at lower $Q^2$,
the uncertainties on \ztl are inflated to 100\% for the region
$0 \leqslant Q^2 < 1$~GeV$^2$ for all $W^2$.
In the numerical calculations the uncertainties on the proton $\gZ$
structure function parametrizations are taken to be the same as those
used in the \bgZv calculation in Ref.~\cite{Hall2013hta}, and a 5\%
uncertainty is assumed for the nucleon elastic contributions.

\begin{table}[b]
\begin{center} 
\caption{Contributions to \regzv\ from Regions~I, II and III,
	and the	total, at the kinematics of the $Q_{\text{weak}}$,
	MOLLER, and MESA experiments.}
\begin{tabular}{cccc}			\hline \hline \vspace*{-0.36cm} \\
	  &  \multicolumn{3}{c}{ \regzv\ ($\times 10^{-3}$)}  \\  \cline{2-4} 
	  & \ \ \qwe \ \   
	  & \ \ \mo \ \
	  & \ \ MESA \ \		\\ 
Region\ \ & ($E=1.165$~GeV)   
	  & ($E=11$~GeV)
	  & ($E=0.18$~GeV)		\\ \hline\vspace*{-0.36cm} \\
I 	  & $4.3 \pm 0.4\ $
	  & $2.5 \pm 0.3$
	  & $1.0 \pm 0.1$		\\
II 	  & $0.4 \pm 0.05$
	  & $3.2 \pm 0.5$
	  & $0.06 \pm 0.01$		\\
III	  & $0.7 \pm 0.04$
	  & $5.5 \pm 0.3$
	  & \ $0.1 \pm 0.01$		\\ \hline
Total	  & $5.4 \pm 0.4\ $
	  & \!\!\!\! $11.2 \pm 0.7$
	  & \!\!\!   $1.2 \pm 0.1$	\\ \hline \hline
\end{tabular}
\label{tab:ReBox}
\end{center}
\end{table}

\begin{figure}[t]
\begin{center}
\includegraphics[width=0.85\textwidth]{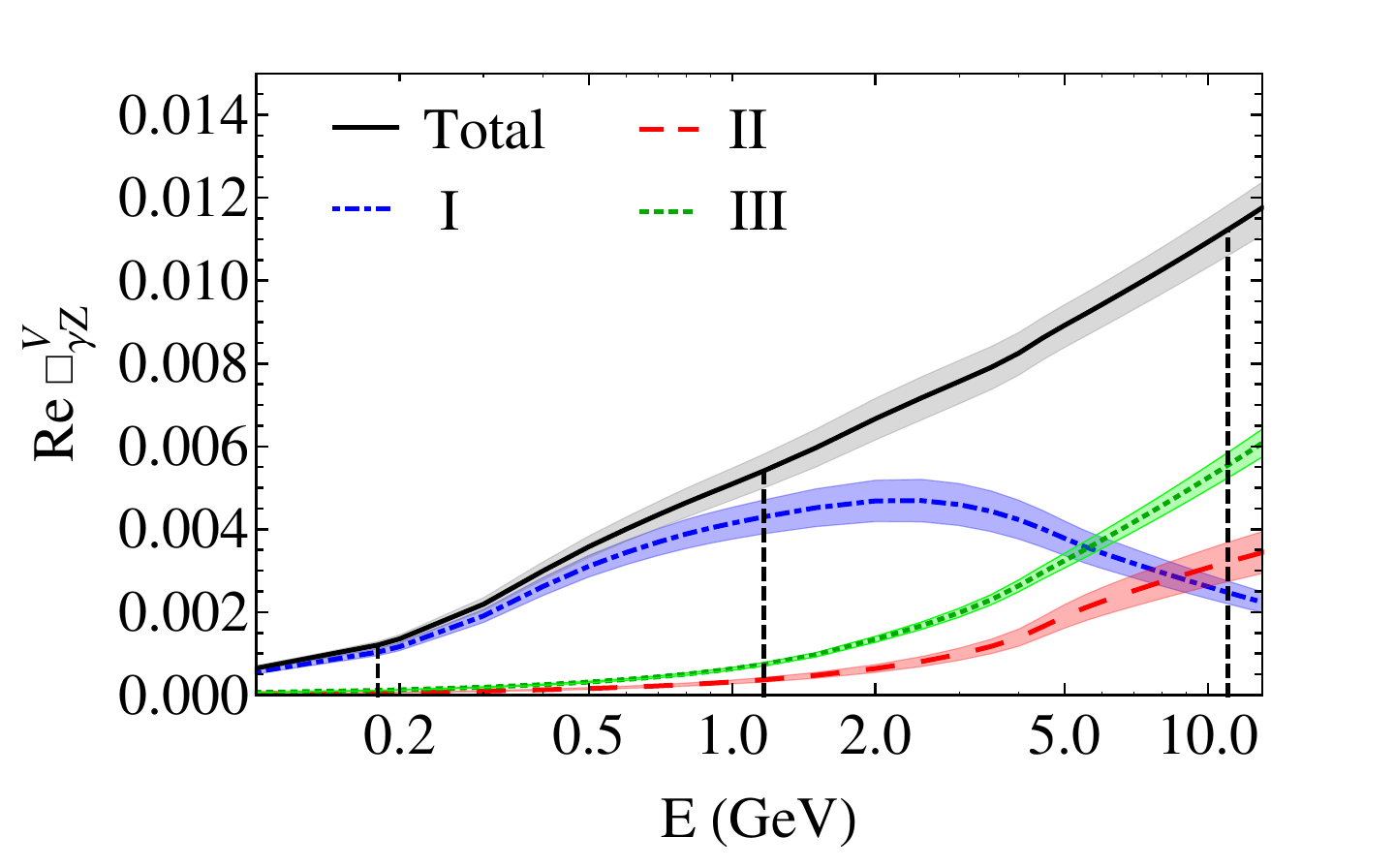}
\caption{Energy dependence of the $\gZ$ box correction,
	$\Re e\, \square_{\gZ}^V$, to $Q_W^p$.
	The contributions from various regions in $W^2$ and $Q^2$
	(Regions I, II and III)	are shown separately, as is the
	total (solid curve).  The dashed vertical lines indicate the
	beam energies of the various parity-violating experiments
	($E = 0.18$~GeV for MESA \cite{MESA},
	 $E = 1.165$~GeV for \qwe \cite{Androic2013}, and
	 $E = 11$~GeV for MOLLER \cite{MOLLER}.}
\label{fig:QRBLoK}
\end{center}
\end{figure}

Using the $\gZ$ structure functions obtained from the newly fitted
\ztl values, the $\Re e\, \square_{\gZ}^V$ correction is displayed
in Fig.~\ref{fig:QRBLoK} as a function of beam energy, with a
breakdown of the individual contributions from different regions
given in Table~\ref{tab:ReBox}.
At the incident beam energy $E = 1.165$~GeV of the \qwe experiment,
the total correction is found to be
\bea
\Re e\, \square_{\gZ}^V &=& (5.4 \pm 0.4) \times 10^{-3}.
\label{eq:ReBox_final}
\eea
This is in good agreement with the value
$\Re e\, \square_{\gZ}^V = (5.57 \pm 0.36) \times 10^{-3}$
found in the previous analysis \cite{Hall2013hta}.
In particular, even though the values of the continuum rotation
parameters in the earlier fit were somewhat different
($\kappa_C^T = 0.65 \pm 0.14$ and $\kappa_C^L = -1.3 \pm 1.7$
with matching to the total DIS structure functions at
$Q^2=2.5$~GeV$^2$), the central value of
$\Re e\, \square_{\gZ}^V$ remains relatively unaffected.

The largest contribution to \regzv\ at the \qwe energy is still
from Region~I, which makes up $\approx 80\%$ of the total,	
with its error dominating the total uncertainty.
Of this, $\approx 2/3$ is from the traditional resonance	
region $W^2 < 4$~GeV$^2$
(of which 61\% is from $Q^2 < 1$~GeV$^2$			
       and 6\% from $Q^2 > 1$~GeV$^2$), and			
$\approx 13\%$ is from $Q^2 < 1$~GeV$^2$ and			
$4 < W^2 < 9$~GeV$^2$.
The contributions from Regions~II and III are $\approx 7\%$	
and $\approx 13\%$, respectively, of the total at the \qwe	
energy, but become more important with increasing energy.
Interestingly, the modified $Q^2$ boundary for Region~III
results in a somewhat smaller contribution from Region~II
($0.4 \times 10^{-3}$ compared with $0.6 \times 10^{-3}$),
while the Region~III contribution has doubled
($0.7 \times 10^{-3}$ compared with $0.35 \times 10^{-3}$)
relative to that in Ref.~\cite{Hall2013hta}.
In effect, moving the $Q^2$ boundary from 2.5~GeV$^2$ to
1~GeV$^2$ shifts $\approx 6\%$ of the total correction		
$\Re e\, \square_{\gZ}^V$ from Regions~I and II to Region~III.

Furthermore, since the $\gamma Z$ structure functions at
$Q^2 < 1$~GeV$^2$ depend on $\kappa_C^{T,L}$, because the $\kappa$
values are refitted at $Q^2 = 1$~GeV$^2$, duality also indirectly
affects the low-$Q^2$ contribution.
Therefore, although duality is formally used only down to
$Q^2 = 1$~GeV$^2$, the constraint influences the $\gamma Z$
calculation below 1~GeV$^2$ as well, as the matching now is
to a more reliable $\gamma Z$ cross section at that point.

While we have assumed the validity of duality for the $F_1^{\gZ}$
and $F_2^{\gZ}$ structure functions down to $Q^2=1$~GeV$^2$, the
possible violations of duality have a minor effect on the analysis.
Even if one takes the maximum violation of duality ($\approx 14\%$)
in the $\gg$ structure functions seen in Fig.~\ref{fig:mugg} at the
lowest $Q^2$ over the entire $1 \leqslant Q^2 \leqslant 2.5$~GeV$^2$
range, the error introduced into the total $\Re e\, \square_{\gZ}^V$
from duality violation is $< 0.1\%$.

Overall, compared with Ref.~\cite{Hall2013hta} the total relative
uncertainty increases marginally, from 6.5\% to 7.4\%, despite the
rather more conservative estimates of the structure function
uncertainty for $Q^2 \lesssim 1$~GeV$^2$ through the inflated
errors on $\kappa_C^{T,L}$.
Note that the same 100\% uncertainties are used in the
transformation of the vector meson dominance model
\cite{Alwall2004wk, Sakurai1972wk} in Region~II.
For Region~III, the LT $F_1^{\gZ}$ and $F_2^{\gZ}$ structure
functions are assigned a 5\% uncertainty for
$Q^2 \geqslant 2.5$~GeV$^2$, which is increased linearly
to 10\% at $Q^2 = 1.0$~GeV$^2$.
%

Since the electromagnetic structure functions are reasonably
well approximated by the LT results even below the traditional
resonance-DIS boundary of $W^2 = 4$~GeV$^2$, we also examine
the effect of lowering the $W^2$ cut into the peripheral
resonance region down to $W^2 = 3$~GeV$^2$.
In this case the contribution from Region~III increases to
$0.9 \times 10^{-3}$, while that from Region~I correspondingly
decreases to $4.2 \times 10^{-3}$, hence leaving the total
essentially unchanged.

At the higher $E=11$~GeV energy of the planned MOLLER experiment
at Jefferson Lab \cite{MOLLER}, the DIS region contributes
about half of the total,
$\Re e\, \square_{\gZ}^V = (11.2 \pm 0.7) \times 10^{-3}$,
with Regions~I and II making up the other 50\%.
This again agrees well with the earlier determination
$\Re e\, \square_{\gZ}^V = (11.5 \pm 0.8) \times 10^{-3}$
from Ref.~\cite{Hall2013loa}.
On the other hand, for the possible future MESA experiment
in Mainz \cite{MESA} at a lower energy, $E=0.18$~GeV,
the bulk of the contribution still comes from Region~I,
but is reduced by a factor of $\sim 4$ compared with the
correction at the \qwe energy.

\section{Conclusion}
\label{sec:Con}

Quark-hadron duality is one of the most remarkable phenomena ever
observed in hadronic physics.
While some aspects of global duality can be formulated in the
language of QCD, such as the relation between the scale independence
of structure function moments and the size of higher twists,
the detailed workings of local duality, for specific regions
of $W^2$ or $x$, are not well understood from first principles.
Nevertheless, there are many marvellous practical applications
to which duality can be put.
For example, the high-energy behavior of hadronic cross sections
can be used to predict averages of resonance properties; and,
conversely, low-$W^2$ data, suitably averaged, can be utilized to
constrain LT parton distributions in difficult to access kinematic
regions.

The latter category appears the most promising approach at present,
with several global PDF analyses \cite{Alekhin2012, CJ13, JR2014}
extending their coverage down to lower $Q^2$ ($Q^2 \gtrsim 1$~GeV$^2$)
and $W^2$ ($W^2 \gtrsim 3$~GeV$^2$) values than in traditional
LT analyses.  This not only increases considerably the available
data base for PDF fitting, it is also one of the few ways currently
available to study PDFs at high $x \sim 1$.

The main implication of duality for the current analysis is the
extension of the LT description of $\gZ$ structure functions to
lower $Q^2$, $Q^2 = 1$~GeV$^2$, than in previous work
\cite{Hall2013hta}.
This serves to reduce the size of the contribution from Region~I,
which has the largest uncertainty associated with the behavior
of the $\gZ$ structure functions at low $Q^2$ and $W^2$.
To account for the possible model dependence of the $\gg \to \gZ$
structure function rotation and the violation of duality at low $Q^2$,
we have assigned rather conservative errors on $F_1^{\gZ}$ and
$F_2^{\gZ}$ in this region.  This is reflected in the increased
uncertainty on this contribution compared with our previous analysis
\cite{Hall2013hta}, which is somewhat offset by the larger
contribution from Region~III that is well constrained by PDFs.

The final result of 
$\Re e\, \square_{\gZ}^V = (5.4 \pm 0.4) \times 10^{-3}$
is consistent with Ref.~\cite{Hall2013hta}, but with a slightly
larger relative uncertainty, which comes almost entirely from
Region~I.  It also agrees with the central value from
Ref.~\cite{Gorchtein2011mz}, although the error there is
$\approx 5$ times larger, which in view of our current analysis
appears to be somewhat overestimated.
Our findings suggest that with the constraints from existing
PVDIS data and PDFs, and now with the further support from
quark-hadron duality, the overall uncertainty in the estimate of
the $\gZ$ box correction is well within the range needed for an
unambiguous extraction of the weak charge from the \qwe experiment.

Further reduction of the uncertainty on the $\gZ$ correction will
come from new measurements of PVDIS asymmetries on the proton,
particularly at the low $Q^2$ and $W^2$ values that are most
relevant at the \qwe energy.  These will also be useful in
constraining the $\gZ$ contribution at the much lower energy
$E=0.18$~GeV of the MESA experiment \cite{MESA}, where we find
the correction to be $\approx 4$ times smaller but even more
dominated by Region~I.
In contrast, for the MOLLER experiment at the higher $E=11$~GeV
energy the dispersion integral is dominated by the DIS region,
which although contributing to a larger overall $\square_{\gZ}^V$
correction, is better determined in terms of PDFs.
These new experiments hold the promise of allowing the most
precise low-energy determination of the weak mixing angle to date,
and providing a unique window on possible new physics beyond the
Standard Model.

\section*{Acknowledgements}

This work was supported by NSERC (Canada), the DOE Contract No. 
DE-AC05-06OR23177, under which Jefferson Science Associates, LLC 
operates Jefferson Lab, and the Australian Research Council through an 
Australian Laureate Fellowship (A.W.T.), a Future Fellowship (R.D.Y.) 
and through the ARC Centre of Excellence for Particle Physics at the 
Terascale.

%
%


\begin{thebibliography}{99}
\bibitem{Nachtmann_note}
Note that the notation for $\mu_1^{(n)}$ and $\mu_2^{(n)}$
in Ref.~\cite{Nachtmann1974} differs from that which we use
in Eqs.~(\ref{eq:mu1}) and (\ref{eq:mu2}).
\end{thebibliography}
\end{document}